\documentclass[12pt, notitlepage]{article}
\usepackage{graphicx}  
\usepackage{amssymb}  
\usepackage{amsmath}  
\usepackage{authblk}
\usepackage{color} % for the demands of the journal
\usepackage{bm} % for the demands of the journal
\begin{document}

\title{The magnetic properties and structure of the quasi-two-dimensional antiferromagnet CoPS$_3$}

\author[1]{A. R. Wildes}
% \ead{wildes@ill.fr}
\author[2]{V. Simonet}
\author[3]{E. Ressouche}
\author[2]{R. Ballou}
\author[4]{G. J. McIntyre}%

\affil[1]{Institut Laue-Langevin, CS 20156, 38042 Grenoble C\'edex 9, France}
\affil[2]{Institut N\'eel, CNRS \& Univ. Grenoble Alpes, 38042 Grenoble, France}
\affil[3]{Univ. Grenoble Alpes, CEA, INAC, MEM, F-38000 Grenoble}
\affil[4]{Australian Centre for Neutron Scattering, Australian Nuclear Science and Technology Organisation, New Illawara Rd, Lucas Heights, NSW 2234, Australia}

\renewcommand\Affilfont{\itshape\small}

\vspace{10pt}
%\begin{indented}
%\item[]v. 1.0, 19.07.17
%\end{indented}
\date{24 June 2017}

\maketitle

\begin{abstract}
The magnetic properties and magnetic structure are presented for CoPS$_3$, a quasi-two-dimensional antiferromagnet on a honeycomb lattice with a N{\'e}el temperature of $T_N \sim 120$ K.  The compound is shown to have XY-like anisotropy in its susceptibility, and the anisotropy is analysed to extract crystal field parameters.  For temperatures between 2 K and 300 K, no phase transitions were observed in the field-dependent magnetization up to 10 Tesla.  Single-crystal neutron diffraction shows that the magnetic propagation vector is  {\bf{k}}= $\left[010\right]$ with the moments mostly along the $\mathbf{a}$ axis and with a small component along the $\mathbf{c}$ axis, which largely verifies the previously-published magnetic structure for this compound.  The magnetic Bragg peak intensity decreases with increasing temperature as a power law with exponent $2\beta = 0.60 \pm 0.01$ for $T > 0.9~T_N$.
\end{abstract}

%\pacs{75.25.$-$j, 75.50.Ee, 75.40.Cx, 75.30.Gw}
%\noindent{\it Keywords}: Layered magnets; antiferromagnetic order; magnetic anisotropy;neutron scattering; magnetometry

%\submitto{\JPCM}
%\maketitle


%Introduction
\section{Introduction}
The MPS$_3$ (M = Mn, Fe, Co, Ni) compounds belong to a family of quasi-two-dimensional materials \cite{Grasso, Brec}.  They are isostructural, with the transition-metal atoms forming a honeycomb lattice in planes that are weakly bound by van der Waals forces.  Their two-dimensional nature gives them remarkable properties, and their electronic structure and intercalation chemistry have been extensively studied.  They are also all antiferromagnetic, and their magnetism is also quasi-two-dimensional.  This has attracted recent attention as it is possible to exfoliate the compounds, and they have potential for magnetic applications in graphene technology \cite{Park}.

They are also model materials for experimental studies on low-dimensional magnetism \cite{Brec}. The magnetic structure and dynamics for some of the compounds, particularly FePS$_3$ \cite{Joy, Lancon} and MnPS$_3$ \cite{Joy, Wildes06}, have been documented and they prove to be very good examples of two-dimensional magnets.  

To date, however, the magnetic properties of CoPS$_3$ have been almost completely overlooked, probably because the compound is the most difficult of the family to synthesize.  They have been the subject of one dedicated article \cite{Ouvrard82}.  The article shows the temperature dependence of the inverse susceptibility of a powdered sample, establishing that the compound is an antiferromagnet with a N{\'e}el temperature of $T_N = 122$ K,  a Weiss temperature of $\theta = -116$ K, and an effective moment of 4.9 $\mu_B$  which is slightly larger than the expected value for a pure spin moment of a $S = 3/2$ Co$^{2+}$ ion implying an orbital contribution.  The article also gives the magnetic structure based on neutron powder diffraction, although neither the data nor the analysis have been presented.  The quoted antiferromagnetic structure consists of ferromagnetic ``zig-zag" chains that are parallel to the $\mathbf{a}$ axis and that are antiferromagnetically coupled along the $\mathbf{b}$ axis, giving CoPS$_3$ a magnetic propagation vector of {\bf{k}}= $\left[010\right]$.  The moments are collinear and lie along the $\mathbf{a}$ axis.  The structure is thus almost identical to that of NiPS$_3$ \cite{Wildes15}.

It is timely to examine the magnetic properties of CoPS$_3$ in more detail.  The renewed interest in this family requires better knowledge of CoPS$_3$ to compare and contrast with its sister compounds.   Furthermore, it may prove to be another useful compound to test model magnetism.  Theory predicts that CoPS$_3$ may show some interesting model behaviour.  The Co$^{2+}$ ions have spin $S = 3/2$ which, on a honeycomb lattice, may lead to a valence-bond ground-state \cite{Affleck}.  Orbital contributions are likely to be important and would probably perturb such a state, however this may also give properties relevant to the  Kitaev-Heisenberg model \cite{Kitaev, Khaliullin}.  Strong anisotropy may lead to Ising or XY-like behaviour in the critical properties.

This article reports a study of the magnetization and magnetic structure of single crystals of CoPS$_3$.  The use of single crystals is important, as previous studies on powdered MPS$_3$ samples have shown issues with partial amorphization of the structure \cite{Wildes15} and have led to misidentification of the magnetic structure \cite{Lancon}.  The results are compared with the other members of the MPS$_3$ family.

% Crystal structure
\section{Crystal structure and domains}
All the MPS$_3$ compounds are monoclinic with space group $C~{2/m}$ and their crystal structures have been measured in detail by Ouvrard \emph{et al.} \cite{Ouvrard85}.  However, as also explained by these authors, the structures are very close to being orthohexagonal and most of the Bragg peaks can equally be indexed using another space group.  The  ambiguity is present because the following relations prove to be a good approximation in these compounds, and particularly so for CoPS$_3$:
\begin{equation}
 \label{eq:acbeta}
 \begin{array}{ll}
 	b &= \sqrt{3}a, \\
	a &= -3c\cos\beta.
 \end{array}
\end{equation}

Diffraction studies are further complicated by the propensity for stacking faults \cite{Ouvrard90} and rotational domains \cite{Murayama} to form in the crystals.  Even the highest-quality crystals will have some of these twinned domains, and care must be taken that magnetic Bragg peaks are indexed using the correct domain.

The domains amount to a three-fold rotation about the $\mathbf{c^{*}}$ axis, which is normal to the $ab$ planes that contain the transition metal atoms \cite{Murayama}.  As shown in the determination of the magnetic structure for FePS$_3$ \cite{Lancon}, the Miller indices for the three are related by the following matrices:
\label{eq:MonoRot}
\begin{equation}
\begin{array}{cl}
\left[ \begin{array}{c} h \\ k \\ l \end{array}\right]^{R1} &
= \left[\begin{array}{rrc} -\frac{1}{2} & \frac{1}{2} & 0 \\ -\frac{3}{2} & -\frac{1}{2} & 0 \\ \frac{1}{2} & -\frac{1}{6} & 1\end{array}\right]\left[ \begin{array}{c} h \\ k \\ l \end{array}\right]^{R2} \\
&\\
& = \left[\begin{array}{rrc} -\frac{1}{2} & -\frac{1}{2} & 0 \\ \frac{3}{2} & -\frac{1}{2} & 0 \\ \frac{1}{2} & \frac{1}{6} & 1\end{array}\right]\left[ \begin{array}{c} h \\ k \\ l \end{array}\right]^{R3},
\end{array}
\end{equation}
where the superscript refers to the relevant domain.

% Experiments
\section{\label{sec:Experiments}Experiments\protect\\ }
% Sample preparation
\subsection{\label{sec:SamplePrep}Sample preparation}
Single-crystal samples of CoPS$_3$ were grown by direct combination of the pure elements, sealed in a quartz ampoule and heated \cite{Grasso,Ouvrard82}.  The ampoule was initially cleaned internally by etching with acid, then rinsing with demineralized water and heating under vacuum to 1000$^\circ$ C for 30 minutes.  Stoichiometric quantities of cobalt, phosphorus and sulphur were placed in the ampoule under an argon atmosphere.  The total mass of the ingredients was 5g, and the purity of the elements was 99.998 \% or better.  The ampoule was then evacuated, sealed under vacuum, and placed in a tube furnace.  The temperature was driven to 100$^\circ$ C in ~10 minutes, then set to ramp to 540$^\circ$ C at a rate of 55$^\circ$ C per 1000 minutes.  Once at temperature, the reaction was left to continue for $\sim$3 months before the furnace was switched off and the ampoule allowed to cool.  A small number of single crystals were found with the characteristic shape, being hexagonal platelets, and  metallic grey colour of CoPS$_3$.   The largest and best of these was $~ 4 \times 2 \times 0.7$ mm$^3$, and was used for the SQUID and high temperature magnetometry and neutron experiments.  Neutron diffraction was used to identify the crystal orientation and to quantify the domain populations, which were found to have the ratios 0.79~:~0.14~:~0.07 and thus the crystal was mostly single-domain.  The magnetization of a second, slightly smaller crystal was measured up to 10 Tesla.
 
% Intro: Magnetometry
\subsection{\label{sec:IntroMag}Magnetometry}
The magnetometry measurements were conducted using magnetometers at the Laboratoire Louis N\'eel, Grenoble. 

A MPMS SQUID magnetometer was used for temperatures between 2 and 350 K.  No glue was used to mount the sample.  The crystal was wrapped in plastic film and then suspended within a plastic drinking straw.  Measurements were taken with the field applied along the $\mathbf{a}$ and $\mathbf{b}$ directions for the majority domain, in the plane of the platelets, and along the $\mathbf{c^{*}}$ direction, corresponding to the normal of the platelet.  The alignment was estimated by eye and not precise, but was probably good to within 10$^\circ$.

The sample magnetization, $M$, was measured as a function of temperature for a number of applied fields, $H$, between 0.01 and 1 Tesla and for both a field-cooled and zero-field-cooled sample.  The magnetization was also measured as a function of applied field up to 5 Tesla at a selection of temperatures.

The purpose-built BS1 magnetometer was used to measure the paramagnetic susceptibility between 300 and 680 K using an axial extraction method.  Again, no glue was used to mount the sample, which was placed in a copper sample-holder whose diamagnetic signal was removed.  Measurements were taken with the field parallel and perpendicular to the $\mathbf{c^{*}}$ direction.  No particular crystallographic axis was chosen for the measurements with the field perpendicular to $\mathbf{c^{*}}$.

The purpose-built BS2 magnetometer was used to measure the magnetization as a function of field up to 10 Tesla at a selection of temperatures using an axial extraction method.  These measurements were performed on a second crystal whose in-plane alignment and domain population had not been characterized.  It was placed in a plexiglas sample holder, again with no glue, whose diamagnetic signal was removed.  Measurements were carried out with the field parallel and perpendicular to $\mathbf{c^{*}}$.

% Intro: Neutron scattering
\subsection{\label{sec:IntroNScat}Neutron Scattering}
All the neutron-scattering measurements were performed on the same crystal used for the SQUID magnetometry measurements.

The nuclear and magnetic scattering were initially characterized with neutron Laue diffraction using the VIVALDI diffractometer at the Institut Laue-Langevin, France \cite{McIntyre}.  The instrument has a cylindrical image-plate detector with a vertical axis.  The incident wavelength band width was 0.8 to 5.2 {\AA}.  The crystal was wrapped in a small amount of aluminium foil before being glued to a pin, thus avoiding contact between the glue and the crystal.  The crystal and mount were then placed in a liquid-helium cryostat before being placed on the instrument and Laue patterns measured as a function of the rotation of the crystal about the cylindrical (vertical) axis.  Measurements were performed between 1.5 and 290 K.

Further  single-crystal diffraction measurements were then performed using the D10 and D23 instruments at the Institut Laue-Langevin.  Both instruments are monochromatic diffractometers.  

D10 was configured with a graphite monochromator.  The wavelength was set to 2.360(2) {\AA}, calibrated against a standard ruby crystal.  Higher-order contamination
was suppressed using a graphite filter.  The crystal was mounted in a liquid-helium flow cryostat on a four-circle Eulerian cradle.  Data were collected using a position-sensitive detector at 4 K and 160 K. The data were corrected for background and integrated using the RACER software \cite{Wilkinson}.  Temperature-dependent measurements of a magnetic Bragg peak were also performed to characterize the critical behaviour of the sub-lattice magnetization.  These latter measurements were performed using a graphite analyser between the sample and detector to improve the signal-to-noise ratio.  The intensities were integrated using a trapezoidal summation, and the background was determined from equivalent measurements above the N{\'e}el temperature.  

D23 was configured with a copper monochromator.  The wavelength was set to 1.280(1) {\AA}.  The crystal was mounted in a liquid-helium cryostat, and data were collected using a single detector at 10 K and 250 K.  The detector can be lifted vertically and rotated horizontally about the sample position, thus accessing sufficient Bragg peaks for a structural refinement.  The data were corrected for background and integrated using the COLL5 software \cite{Coll5}.

After integration, the FullPROF suite \cite{FullPROF} was used to merge and analyse the data from both D10 and D23.  The residuals, as defined in the FullPROF manual and release notes, were calculated using a SHELX-like weighting scheme (Iwg = 3) with default values.  Care was taken to quantify and account for the domain population in the analysis.

% Results
\section{\label{sec:Results}Results}
% Results: Magnetometry
\subsection{\label{sec:ResMag}Magnetometry}
\begin{figure}
  \includegraphics[scale=0.6]{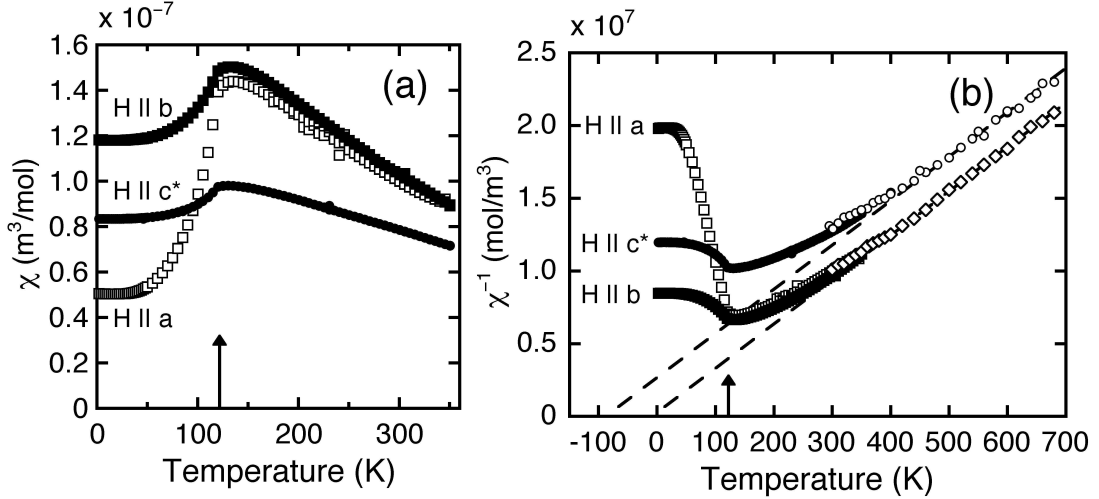}
  \caption{\label{fig:MvT}The (a) susceptibility; and (b) inverse susceptibility for a single crystal of CoPS$_3$.  The field was applied along different directions for the measurements, which are marked in the figure.  Data between 2 and 350 K were measured using a MPMS SQUID magnetometer with an applied field of 0.1 Tesla, while the BS1 magnetometer was used to collect data between 300 and 680 K.   Fits of a straight line to the high-temperature inverse susceptibility data for H parallel and perpendicular to $\mathbf{c^{*}}$ are shown.  The arrows mark the expected N{\'e}el temperature of 122 K.}
\end{figure}

In the linear regime, the susceptibility is equivalent to $M/H$.  Figures \ref{fig:MvT} show this susceptibility and its inverse for CoPS$_3$ as a function of temperature, measured with a field applied along different crystallographic directions.  The data up to 350 K were measured using the MPMS magnetometer with an applied field of 0.1 Tesla.  The higher-temperature measurements, shown as open symbols starting from 295 K, were measured using the BS1 magnetometer.  The data from the two magnetometers agree in the temperature regions where they overlap.

The data in figure \ref{fig:MvT} confirm the previous conclusions that CoPS$_3$ is an antiferromagnet \cite{Ouvrard82}.  All the data show a sharp kink at the expected N{\'e}el temperature of 122 K, and this temperature is shown as an arrow in the figure.  The susceptibility below $T_N$ varies little with temperature for fields applied along $\mathbf{b}$ and $\mathbf{c^{*}}$, suggesting that the ordered moments  are mostly perpendicular to these directions.  The susceptibility decreases more dramatically with decreasing temperature for fields applied along $\mathbf{a}$, although it does not fall to zero and asymptotes to $\approx 5 \times 10^{-8}$ m$^3\cdot$mol$^{-1}$ at the lowest temperatures.  This is consistent with the moments pointing largely along $\mathbf{a}$, as expected \cite{Ouvrard82}.

% Figure 1
% end Figure 1

The susceptibility above $T_N$ is anisotropic, with the susceptibility for fields in the $ab$ plane being larger than that with the field applied along $\mathbf{c^{*}}$, normal to the plane. This indicates an in-plane anisotropy, hinting at XY-like characteristics in the Hamiltonian. The susceptibility with the field applied along $\mathbf{b}$ is slightly larger than for the field applied along $\mathbf{a}$, indicating a weak additional anisotropy within the $ab$-plane. There is a slight hump in the susceptibility above $T_N$, giving a maximum at $T \approx 130$ K, which may be due to two-dimensional critical fluctuations.  A similar hump is seen in the other members of the MPS$_3$ family \cite{Joy, Wildes15}, although it appears to be smallest in CoPS$_3$.  Measurements of field-cooled and zero-field-cooled susceptibilities showed no significant differences.

The inverse susceptibilities for $T \ge 425 K$ were fitted with straight lines, and the fits are also shown in figure \ref{fig:MvT}b.  The gradients are inversely proportional to the Curie constant, $C$.  They were constrained to be equal in the fits, giving a value of $C = \left(3.29 \pm 0.03\right)\times 10^{-5}$ m$^3\cdot$mol$^{-1}\cdot$K.  The Curie constant gives an effective moment of $\mu_{eff} = 4.55~\mu_B$ which is close to the previously quoted value of 4.9 $\mu_B$ \cite{Ouvrard82},.  This differs from a spin-only effective moment for $S = 3/2$ of 3.87 $\mu_B$, suggesting that the orbital momentum is not completely quenched.  The x-intercepts were found to be  $-87.9$ K for $\mathbf{H}~||~\mathbf{c^{*}}$ and $-9.2$ K for  $\mathbf{H}~\perp~\mathbf{c^{*}}$.

The observed anisotropies at high temperatures ($\gtrsim 400$ K) may be further analysed if it is assumed that they are due to crystalline electric fields.  The effects of these fields can be approximated by using first-order perturbation theory to consider the zero-field splitting of a spin $S=3/2$ multiplet, corresponding to orbitally-quenched Co$^{2+}$ ions.  The magnitude of the Curie constant suggests that the Co$^{2+}$ ions are not fully quenched.  However the estimated effective orbital contribution is $L \approx 0.32$, which is small enough that the approximations used for first-order perturbation theory are still valid.

The potential for the crystalline electric field, $V$, may be written as a function of the crystal field parameters, $A^q_k$, and Stevens operators, $\tilde{O}^q_k$, as:
\begin{equation}
	V=\sum_{k}\sum^k_{q=-k}A^q_{k\alpha}\tilde O^q_{k\alpha}
\end{equation}
The magnetization $M_\alpha$ is measured when a weak magnetic field $H_{\alpha}$ is applied parallel to an axis $\bm{\alpha}$ of the crystal.  By using perturbation theory to first order, $M_\alpha$ can be expanded at high temperature to give \cite{Boutron1969}:
\begin{equation}
M_{\alpha}=\frac{CH_{\alpha}}{T}\left(1-\frac{1}{k_BT}\frac{\text{Tr}[V\tilde{O}^0_{2\alpha}(S)]}{S(S+1)(2S+1)}\right)+O({1\over T^3})
\label{eq:M_Steven}
\end{equation}
where $C=g^2_J\mu_B^2S(S+1)/3k_B$ is the Curie constant.  The Stevens operator $\tilde{O}^0_{2\alpha}(S) = 3 S_z^2-S(S+1)$ when the quantization axis $\mathbf{z}$ is parallel to the axis $\bm{\alpha}$.  

The potential, $V$, can be given as a function of the appropriate Stevens operators $O^q_k$ for the spin $S=3/2$ multiplet.  The operators must account for the triangular constraint $0 \le k \le 2S = 3$, for the parity symmetry, and for the two-fold symmetry associated with the $4g$ Wyckoff site occupied by the Co$^{2+}$ ions in CoPS$_3$.  The local two-fold axis of the $4g$ site is oriented along $\mathbf{b}$.  Defining this axis as the quantization axis $\mathbf{z}$, the potential $V$ can then be written:
\begin{equation}
	V = A_2^0 \tilde{O}^0_2+A_2^2(c) \tilde{O}^2_2(c)+A_2^2(s) \tilde{O}_2^2(s)
\label{eq:V_Steven}
\end{equation}
with $\tilde{O}^0_{2} = 3 S_z^2-S(S+1)$, $\tilde{O}^2_2(c) = S_x^2-S_y^2$ and $\tilde{O}_2^2(s) = S_xS_y+S_yS_x$.

Equation \ref{eq:M_Steven} can be simplified using equation \ref{eq:V_Steven} and the identity:
\begin{equation}
	\text{Tr}[\tilde{O}^q_{k\alpha}(S) \tilde{O}^0_{2\alpha}(S)]=\frac{1}{5}S(S+1)(2S+1)(2S-1)(2S+3)\delta_{k,2}\delta_{q,0}.
\end{equation}
If the exchange interactions are considered to be isotropic and are treated in the molecular field approximation, the high-temperature inverse susceptibilities along the $\mathbf{b}$ axis, $\chi_b$, and along an axis $\mathbf{u}$ in the $ac$ plane, $\chi_{u}$, are given by:
\begin{eqnarray}
\label{eq:inversechib}
\frac{1}{\chi_b} & = & \frac{1}{C} \left\{T-\Theta + \Lambda A_2^0\right\} \\
\label{eq:inversechiu}
\frac{1}{\chi_{u}} & = & \frac{1}{C} \left\{T-\Theta + \Lambda \left(-\frac{A_2^0}{2}+\frac{A_2^2(c) \cos2\phi}{2}-\frac{A_2^2(s) \sin2\phi}{2}\right)\right\}
\end{eqnarray}
where $\Lambda = (2S-1)(2S+3)/5k_B$, $\Theta$ is the Curie-Weiss temperature, and $\phi$ is the angle between the axis $\mathbf{u}$ and a selected axis in the $ac$ plane.  Selecting axis $\mathbf{a}$ results in $\phi = 0$ for $\chi_a$ and $\phi = 3\pi/2$ for $\chi_{c^*}$.  Substituting these values into equation \ref{eq:inversechiu} results in the expressions for the inverse susceptibilities along $\mathbf{a}$ and $\mathbf{c^{*}}$:
\begin{eqnarray}
\label{eq:inversechia}
\frac{1}{\chi_{a}} & =  \frac{1}{C} \left\{T-\Theta + \Lambda \left(-\frac{A_2^0}{2}+\frac{A_2^2(c)}{2}\right)\right\} \\
\label{eq:inversechicstar}
\frac{1}{\chi_{c^{*}}} & =  \frac{1}{C} \left\{T-\Theta + \Lambda \left(-\frac{A_2^0}{2}-\frac{A_2^2(c)}{2}\right)\right\}
\end{eqnarray}

Estimates for the crystal field parameters may now be derived from the linear fits to the high-temperature data in figure \ref{fig:MvT}b.   The susceptibilities  $\chi_a (T)  \approx \chi_b (T)$ at high temperature, hence $A_2^2(c) \approx 3 A_2^0$ from equations \ref{eq:inversechib} and \ref{eq:inversechia}.  The x-intercepts from the linear fits give $\chi_b (T+87.9)  \approx  \chi_{c^*} (T+9.2)$ at high temperatures, giving  $3\Lambda A_2^0 = -78.7$ K from equations  \ref{eq:inversechib} and \ref{eq:inversechicstar}.  Substituting these values gives  $A_2^0 = -10.9$ K and $A_2^2(c) = -32.7$ K.

The result that  $A^2_2\left(c\right) < 0$ sets $\mathbf{a}$ to be the easy axis in the $ac$ plane.  Furthermore, the result that $\left|A^2_2\left(c\right)\right| > \left|A^0_2\right|$ means that it is less easy to magnetize along the $\mathbf{b}$ direction than along $\mathbf{a}$.  Both of these results are consistent with what is observed experimentally.

A Curie-Weiss temperature of $\theta = -35.4$ K is given by substituting $\Lambda A^0_2$ and $1/\chi_b = 0$ at $T = -9.2$ K into equation \ref{eq:inversechib}.  This is also equal to the weighted mean of the x-intercepts.  The value for $\theta$ is smaller than the previously determined value of $-116$ K from a powder sample \cite{Ouvrard82} which was probably overestimated due to the use of a more limited and lower temperature range of 300 K $\le T \le$ 420 K.

A value for $A^2_2\left(s\right)$ cannot be explicitly determined without a magnetization measurement along a third direction in the $ac$ plane.  However, if the quantization axis $\mathbf{z}$ is rotated from $\mathbf{b}$ to $\mathbf{c^{*}}$ equation \ref{eq:V_Steven} becomes:
\begin{equation}
	V = -(A_2^0/2+A_2^2(c)/2) O_2^0 + (-3A_2^0/2+A_2^2(c)/2) O_2^2(c) - A_2^2(s) O_2^1(s).
\label{eq:Vrot}
\end{equation}
On substituting the previously-determined result $A_2^2(c) \approx 3 A_2^0$, this becomes:
\begin{equation}
	V = -2A_2^0 O_2^0 - A_2^2(s) O_2^1(s).
\end{equation}
Noting that $A^0_2 < 0$,  the hard axis is $\mathbf{c^{*}}$ if $A^2_2\left(s\right)$ is vanishingly small.  This is consistent with the proximity of the structure to an orthohexagonal space group, which singularizes the $\mathbf{c^{*}}$ axis, and hence $A^2_2\left(s\right)$ is likely to be small.

\begin{figure}
  \includegraphics[scale=0.6]{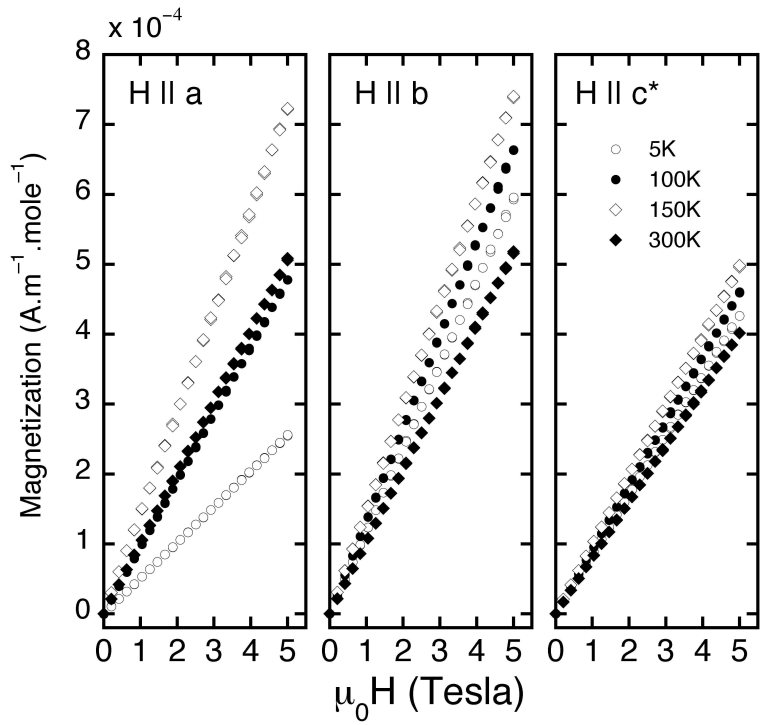}
  \caption{ \label{fig:MvHsquid}Magnetization as a function of applied field along different crystallographic directions.  The measurements were performed using the MPMS magnetometer.}
\end{figure}

\begin{figure}
  \includegraphics[scale=0.6]{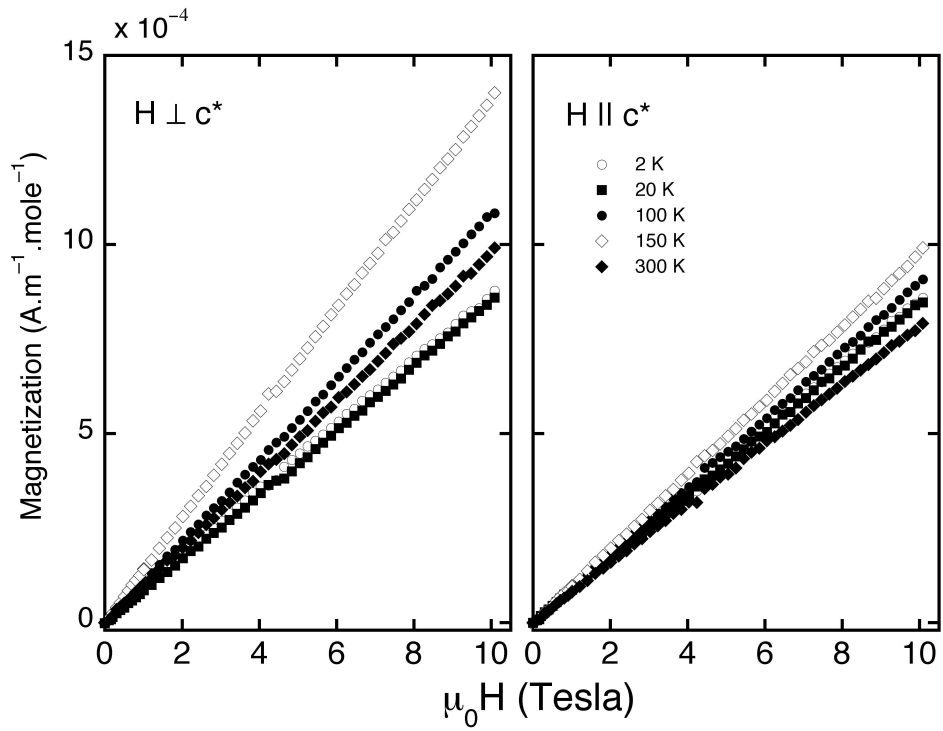}
  \caption{\label{fig:MvHBS2}Magnetization as a function of applied field for the field parallel and perpendicular to the $\mathbf{c^{*}}$ axis.  The measurements were performed using the BS2 magnetometer.}
\end{figure}

Measurements of the magnetization as a function of applied field at a selection of temperatures are shown in figures \ref{fig:MvHsquid} and \ref{fig:MvHBS2}.  Figure \ref{fig:MvHsquid} shows the results from the MPMS magnetometer, while figure \ref{fig:MvHBS2} shows the results from the BS2 magnetometer.  The data for $\mathbf{H}~||~\mathbf{c^{*}}$ match well between the two magnetometers.  The measurements were performed on two different crystals.  The $\mathbf{H}~\perp~\mathbf{ c^{*}}$ data from the BS2 magnetometer fall between the $\mathbf{H}~||~\mathbf{a}$ and $\mathbf{H}~||~\mathbf{b}$ data for the MPMS magnetometer, however this is unsurprising as, unlike the crystal measured on the MPMS magnetometer, the domain population was not verified for the crystal measured on BS2.  The $\mathbf{H}~\perp~\mathbf{c^{*}}$ data in figure \ref{fig:MvHBS2} thus represent a weighted average over the domain population in this crystal.  All of the data show a straight-forward linear increase of magnetization with field at all temperatures, with no strong evidence for field-induced phase transitions up to 10 Tesla.  

% Figure 2
% end Figure 2

% Figure 3
% end Figure 3

% Results: Neutron scattering
\subsection{\label{sec:ResNScat}Neutron Scattering}

\begin{figure}
  \includegraphics[scale=0.6]{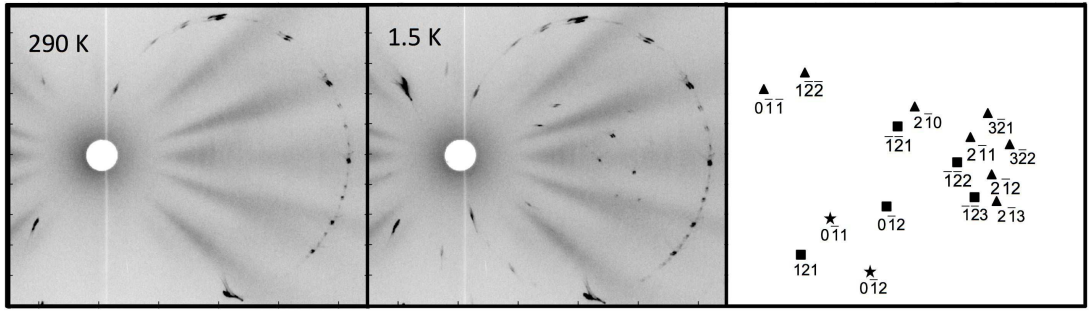}
  \caption{\label{fig:Laue} Neutron Laue diffraction patterns of CoPS$_3$ at 290 K (left panel) and 1.5 K (middle panel). The new Laue spots at 1.5 K are due to the onset of long-ranged antiferromagnetic order.  The Miller indices for the new spots are shown in the right panel.  The different symbols (triangle, square, star) in the right panel represent the crystallographic domains used to index the spots.}
\end{figure}

Preliminary measurements carried out on the VIVALDI Laue diffractometer showed Laue spots that could be indexed using LAUEGEN in the CCP4 software suite \cite{Campbell}.  Figure \ref{fig:Laue} shows parts of representative patterns.  The left-hand panel shows a pattern at 290 K, where only spots due to nuclear Bragg peaks appear.  These spots could be all indexed with the expected space group of $C~{2/m}$ and lattice parameters for CoPS$_3$ \cite{Ouvrard85}.  Close inspection shows that the spots have irregular shapes and  a large degree of streaking.  This is due to the two-dimensional nature of the crystals and, unfortunately, meant that it was not possible to determine accurately the integrated intensities required for a detailed structure refinement.

% Figure 4
% end Figure 4

However, the patterns can be indexed and used to determine the magnetic propagation vector below $T_N$.  Inspection of the pattern at 1.5 K shows a number of new spots due to the onset of antiferromagnetic order.  The spots could all be indexed and attributed to one of the three crystallographic domains given by equations \ref{eq:MonoRot}.  The new spots with their indices are marked in the right-hand panel of figure \ref{fig:Laue}.  The symbol used to mark the spot also indicates which crystallographic domain was used for its index, with the three domains indicated respectively by stars, triangles and squares.

All the Laue spots could be indexed with integer Miller indices, indicating that the magnetic unit cell is the same size as the nuclear unit cell and there is no incommensurate order.  The indices are all forbidden in the $C~{2/m}$ space group, hence the magnetic structure does not have {\bf{k}} = 0.  

\begin{figure}
  \includegraphics[scale=0.6]{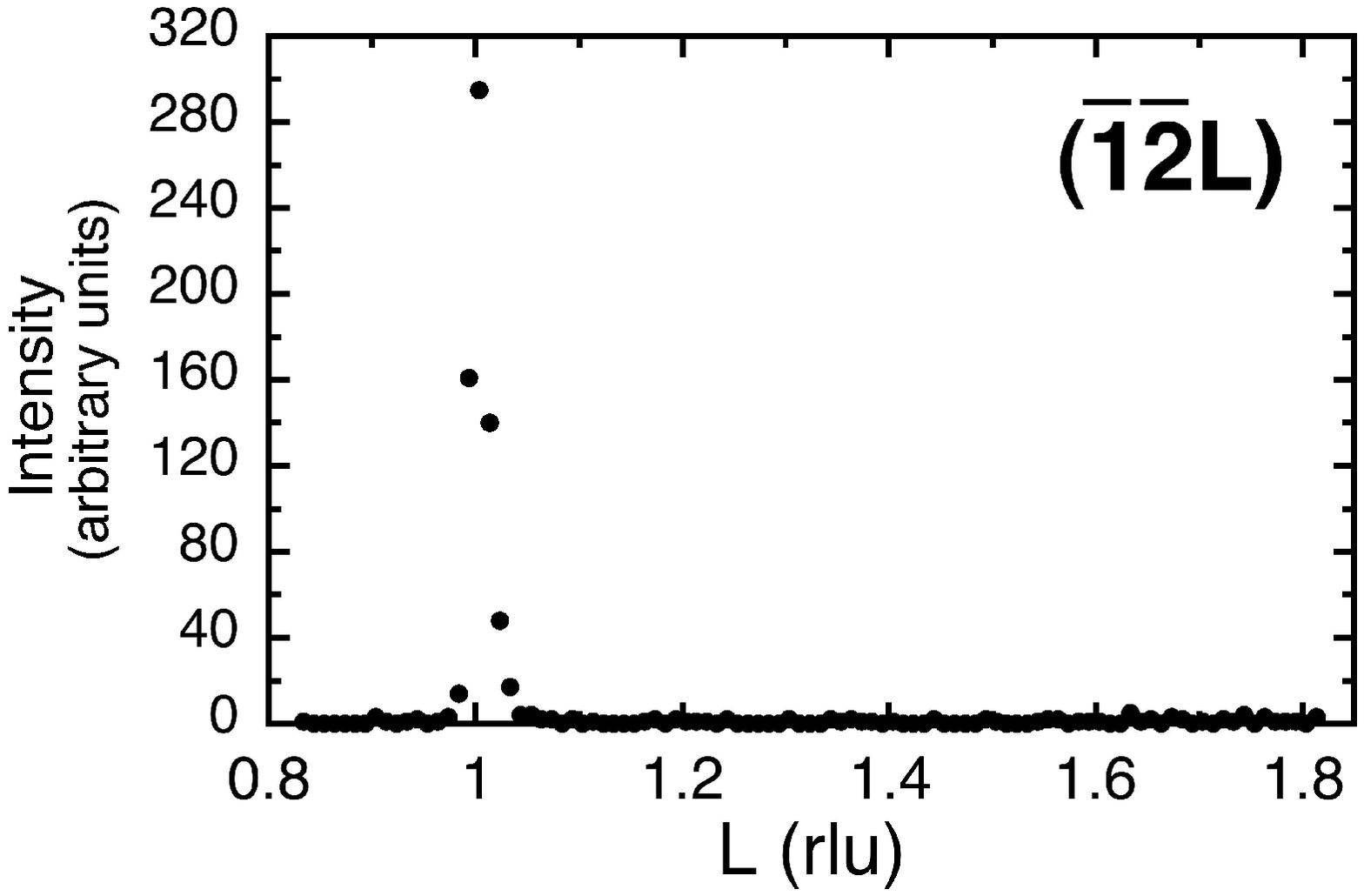}
  \caption{\label{fig:Lscan} Neutron diffraction measurement from D10 along $\left(\overline{1}\overline{2}l\right)$ of CoPS$_3$ at 4 K.}
\end{figure}

The magnetic propagation vector was previously established to be {\bf{k}}= $\left[010\right]$ \cite{Ouvrard82} and measurements on D10 and D23 showed this to be correct.  Figure \ref{fig:Lscan} shows a scan along $\left(\overline{1}~\overline{2}~l\right)$ in the dominant domain at 4 K on D10.  A clear magnetic Bragg peak is visible at $\left(\overline{1}~\overline{2}~1\right)$, consistent with the expected propagation vector.  Scans along other reciprocal lattice trajectories were also consistent with this propagation vector, and measurements of individual peaks showed that the magnetic Bragg peaks were strongest along $\left(0k0\right)$.  The Bragg peaks consistent with {\bf{k}}= $\left[010\right]$ were subsequently measured on D10 and D23 and the crystallographic and magnetic structures refined.

% Figure 5
% end Figure 5

\begin{figure}
  \includegraphics[scale=0.6]{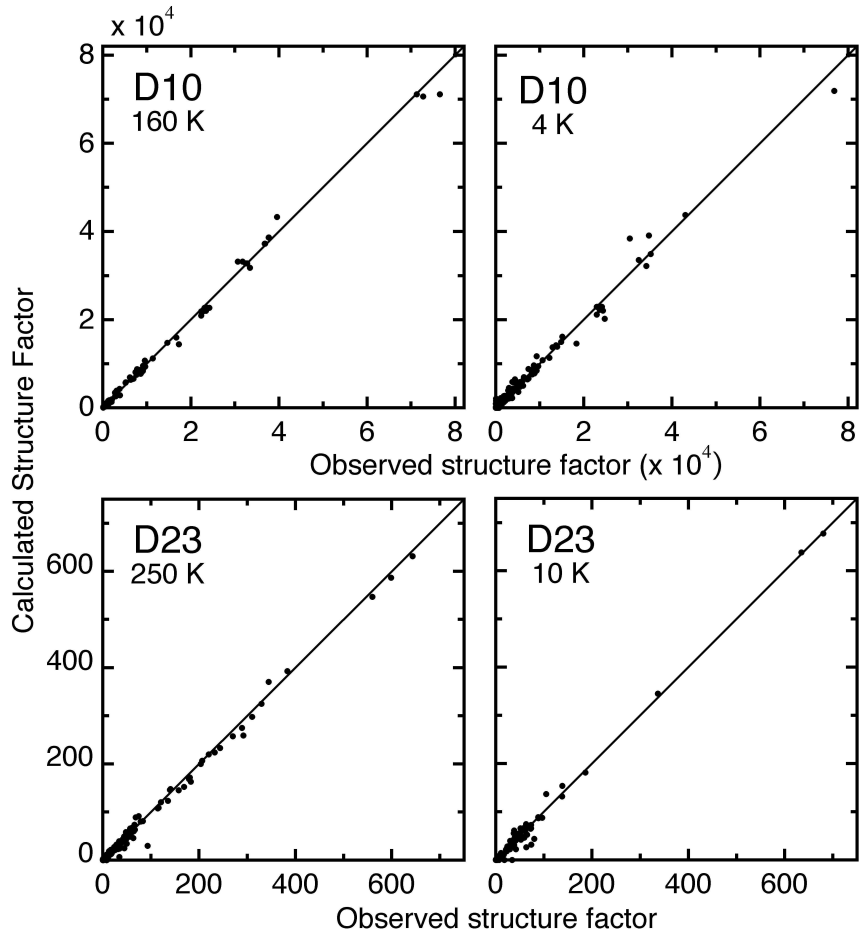}
  \caption{\label{fig:SingXtal} Observed and calculated structure factors from refinements performed on data measured using D10 and D23.  The left-hand panels show refinements of purely the nuclear structure, using data measured above the N{\'e}el temperature.  The right-hand panels show refinements of the nuclear and magnetic structures from data measured below $T_N$.  The results of the nuclear refinement are shown in table \ref{tab:Nuc} and of the magnetic structure are shown in table \ref{tab:Mu}.}
\end{figure}

The data from D10 and D23 were refined separately to check for consistency.  The DataRed program within the FullPROF suite was used to merge the data, using equations \ref{eq:MonoRot} to account for crystallographic domains.  The crystallographic structure was then refined from data measured above the N{\'e}el temperature.  The refined structure parameters and residuals are shown in table \ref{tab:Nuc}, and the left-hand panels of figure \ref{fig:SingXtal} show the comparison between the observed and calculated structure factors for the measured peaks.  The refinements are satisfactory for both data sets, which give parameters that largely agree both between data sets and with the  previously published structure \cite{Ouvrard85}.  The only major difference is found in the isotropic temperature factors, $B_{iso}$, for the lower-temperature D10 data, which had to be set to be equivalent to have a stable refinement.

% Table 1
\begin{table}
\caption{\label{tab:Nuc}The refined crystal-structure parameters for CoPS$_3$ for the separate measurements on D10 and D23 at 160 K and 250 K respectively.  The space group was $C~2/m$.  The atoms are marked with their Wyckoff positions.  The fractional coordinates are given with respect to the monoclinic unit cell. }
%\begin{indented}
%\item[]\begin{tabular}{@{}lccccc}
\begin{tabular}{lccccc}
\hline
\hline
D10 \\
Reflections: & Scanned: & 237 \\
                      & Independent: & 125 \\
 \hline
& $a = 5.895\left(2\right)$ {\AA} & $b = 10.19\left(1\right)$ {\AA} & $c = 6.603\left(3\right)$ {\AA} & $\beta = 107.18\left(3\right)$ \\
& $R_{F^2} = 4.983$ & $R_{wF^2} = 16.40$ & $R_F = 3.893$ & $\chi^2 = 0.599$ \\
\hline
Atom                                      & $x$                             & $y$                              & $z$                             & $B_{iso}$                 \\
\hline
Co $\left(4g\right)$              & 0                                  & $0.3314\left(9\right)$ & 0                                 & $0.21\left(1\right)$ \\
P $\left(4i\right)$                  & $0.057\left(1\right)$ & 0                                  & $0.1690\left(8\right)$ & $0.21\left(1\right)$ \\
S $\left(4i\right)$                  & $0.746\left(2\right)$ & 0                                  & $0.244\left(1\right)$ & $0.21\left(1\right)$ \\
S $\left(8j\right)$                  & $0.252\left(2\right)$ & $0.1690\left(5\right)$ & $0.2470\left(8\right)$ & $0.21\left(1\right)$ \\
\\
\hline
\hline
D23 \\
Reflections: & Scanned: & 592 \\
                      & Independent: & 225 \\
\hline
& $a = 5.897\left(2\right)$ {\AA} & $b = 10.216\left(4\right)$ {\AA} & $c = 6.664\left(3\right)$ {\AA} & $\beta = 107.17\left(3\right)$ \\
& $R_{F^2} = 7.596$ & $R_{wF^2} = 14.64$ & $R_F = 6.508$ & $\chi^2 = 0.3345$ \\
\hline
Atom                                      & $x$                             & $y$                              & $z$                             & $B_{iso}$                 \\
\hline
Co $\left(4g\right)$              & 0                                  & $0.3323\left(5\right)$ & 0                                 & $0.8\left(1\right)$ \\
P $\left(4i\right)$                  & $0.0574\left(7\right)$ & 0                                  & $0.1710\left(7\right)$ & $0.49\left(9\right)$ \\
S $\left(4i\right)$                  & $0.746\left(2\right)$ & 0                                  & $0.246\left(1\right)$ & $0.7\left(1\right)$ \\
S $\left(8j\right)$                  & $0.250\left(1\right)$ & $0.1679\left(3\right)$ & $0.245\left(1\right)$ & $0.6\left(1\right)$ \\
\hline
\hline
\end{tabular}
%\end{indented}
\end{table}
% End table 1

The previous study suggested that CoPS$_3$ had some site disorder, amounting to 3{\%}, in the Co and P positions \cite{Ouvrard85}.  Including site disorder did not substantially  improve the quality of the refinement, which depended far more on the presence of crystallographic domains.  Hence, for simplicity, the site disorder was ignored in the subsequent magnetic refinement.

% Figure 6
% end Figure 6

The irreducible representations of the groups compatible with {\bf{k}}= $\left[010\right]$ in the $C~{2/m}$ space group were determined using the BasIreps program within the FullProf suite.  They are summarized in table \ref{tab:Ireps}.  The correct group was immediately obvious from inspection of the Bragg peak intensities.  Ireps $\left(1\right)$ and $\left(3\right)$ could be rejected because they give magnetic Bragg peaks at $\left(100\right)$ and $\left(030\right)$, where no magnetic intensity was observed.  The strongest Bragg peak was found at $\left(010\right)$, which is not compatible with Irep$\left(2\right)$.  The magnetic structure was therefore refined from the low-temperature data using Irep$\left(4\right)$.  The comparison between the calculated and observed structure factors is shown in the right-hand panels of figure \ref{fig:SingXtal}.

% Table 2
\begin{table*}
\caption{\label{tab:Ireps}The moment components and symmetry operators associated with the four possible irreducible representations for a magnetic propagation vector of $\left[010\right]$ in combination with the space group $C~{2/m}$.}
%\begin{indented}
%\item[]\begin{tabular}{@{}lccccc}
\begin{tabular}{lccccc}
\hline
\hline
 	                    			&						& IRep(1) 				& IRep(2)				& IRep(3) 				& IRep(4) \\
\hline
Moments: \\ 
& $\left(x,y,z\right)$  							& $\left(0,M_y,0\right)$ 	& $\left(0, M_y,0\right)$	& $\left(M_x,0,M_z\right)$	& $\left( M_x,0, M_z\right)$ \\
& $\left(\overline{x},1+\overline{y},\overline{z}\right)$	& $\left(0,M_y,0\right)$	& $\left(0,-M_y,0\right)$	& $\left(M_x,0,M_z\right)$	& $\left(-M_x,0,-M_z\right)$ \\
\hline
Symmetries: \\
&  ~~1: $\left(0,0,0\right)$						&	1				&	~~1				&	~~1				&	~~1		\\
& ~~2: $\left(0,y,0\right)$						&	1				&	~~1				&	$-1$				&	$-1$		\\
&  $-1$: $\left(0,0,0\right)$						&	1				&	$-1$				&	~~1				&	$-1$		\\
& ~~m: $\left(x,0,z\right)$						&	1				&	$-1$				&	$-1$				&	~~1		\\
\hline
\end{tabular}
%\end{indented}
\end{table*}
%End table 2

The results of the refinement, allowing the moments to have components along the $\mathbf{a}$ and $\mathbf{c}$ axes, are shown in table \ref{tab:Mu} along with the residuals. The quality of the refinements are satisfactory and are coherent and consistent between the D10 and D23 data sets. The moments lie mostly along $\mathbf{a}$ with a small out-of-plane component. The ordered moments are slightly greater than 3 $\mu_B$ (the spin-only moment on a Co$^{2+}$ ion) in agreement with the paramagnetic moment derived from fitting the inverse susceptibility at high temperature.

The previously published magnetic structure for CoPS$_3$ is consistent with Irep$\left(4\right)$.  It stated that the moments lay only along the $\mathbf{a}$ axis.  Refinements with this constraint were also performed, and the results are also shown in table \ref{tab:Mu}.  These refinements are poorer, hence an out-of-plane component is most likely present.  

\begin{figure}
  \includegraphics[scale=0.6]{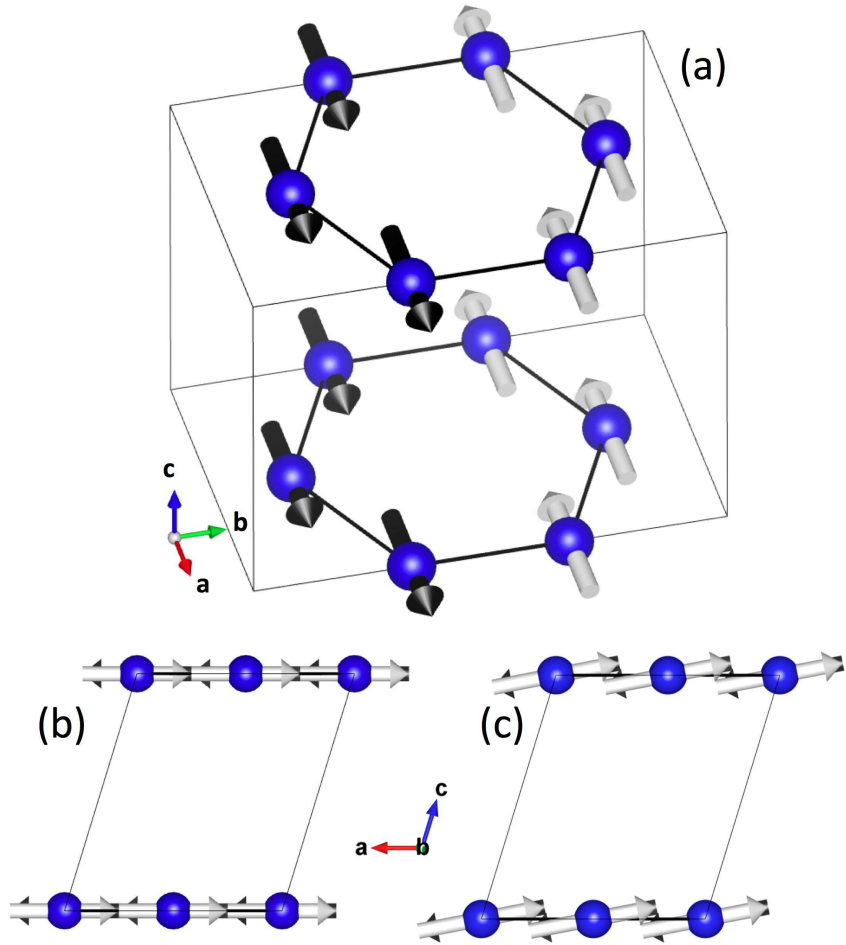}
  \caption{\label{fig:CoPSMag} Magnetic structure for CoPS$_3$.  (a) Isometric view of the unit cell. (b) View along the $b$-axis for the previously-published structure \cite{Ouvrard82}, showing the moments lying along the $a$ axis. (c) View along the $b$ axis for the refined structure in the present work, showing a small out-of-plane component to the moments.}
\end{figure}

A schematic of the magnetic structure for CoPS$_3$ is shown in figure \ref{fig:CoPSMag}.  The originally published structure is shown in figure \ref{fig:CoPSMag}(b), while the results for the current refinement are shown in figure \ref{fig:CoPSMag}(c).  The current refinement results in a structure that is qualitatively identical to that for NiPS$_3$ \cite{Wildes15}.

% Table 3
\begin{table*}
	\caption{\label{tab:Mu} Ordered moments from refining the low-temperature data from D10 and D23 using Irep$\left(4\right)$.  Two refinements were performed: one constraining the moments to lie only along $\mathbf{a}$; and a second with the moments free in the $\left(a,c\right)$ plane.  The residuals for each refinement are also shown.  All moment values are given in $\mu_B$.  Note that both nuclear and magnetic Bragg peaks were measured and that the refinements included both nuclear and magnetic structures.}
%	\begin{indented}
%	\item[]\begin{tabular}{@{}l cc cc}
	\begin{tabular}{l cc cc}
\hline
\hline
 & \multicolumn{2}{c}{D10} & \multicolumn{2}{c}{D23} \\
\hline
Scanned reflections: & \multicolumn{2}{c}{421} & \multicolumn{2}{c}{269}\\
Independent reflections: & \multicolumn{2}{c}{417} & \multicolumn{2}{c}{121}\\
\hline
Irep$\left(4\right)$& $\left( M_x,0, M_z\right)$ &  $\left( M_x,0, 0\right)$ & $\left( M_x,0, M_z\right)$ &  $\left( M_x,0, 0\right)$\\
\hline
Moment along $a$ &  $3.13\left(8\right)$       &   $3.33\left(7\right)$ &  $3.3\left(1\right)$         &   $3.62\left(6\right)$\\
Moment along $c$ & $-0.6\left(1\right)$        &   $-$                             &  $-0.6\left(1\right)$       &   $-$ \\
Total moment          &  $3.36\left(9\right)$     &  $3.33\left(7\right)$   &  $3.5\left(1\right)$        &   $3.62\left(6\right)$\\
$R_{F^2}$               &  8.759                              &   9.14                          &  10.50                             &   11.51 \\
$R_{wF^2}$            &  46.47                              &   47.30                       &  22.77                             &   25.06  \\
$R_F$                      &  11.32                              &   11.71                        & 9.714                             &   10.56  \\
$\chi^2$                   &  4.198                              &   4.359                         & 0.8128                          &   0.9853 \\
\hline
 	\end{tabular}
%	\end{indented}
\end{table*}
% End table 3

% Figure 7
% end Figure 7

\begin{figure}
  \includegraphics[scale=0.6]{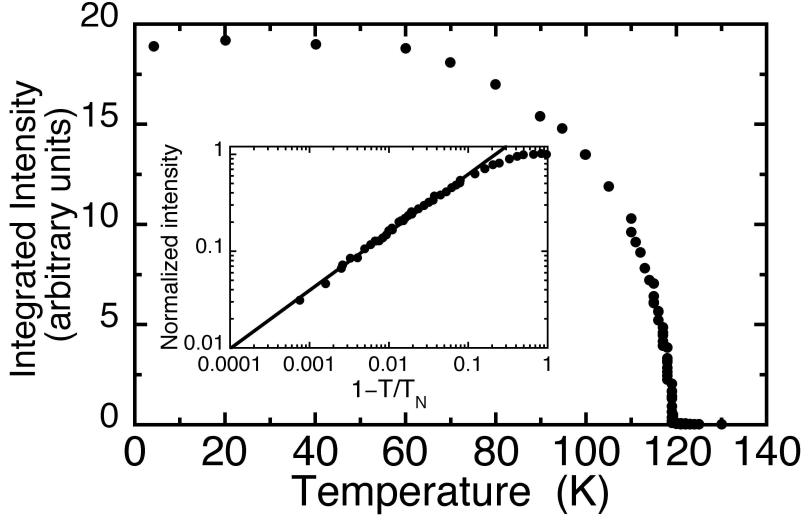}
  \caption{\label{fig:beta}  Integrated intensity of the $\left(010\right)$ Bragg peak as a function of temperature.  The N{\'e}el temperature was established to be $119.1\pm0.1$ K.  The insert shows the normalized data plotted against reduced temperature with a fit to a power law.  The fitted exponent was found to be $2\beta = 0.60\pm 0.01$.}
\end{figure}

The temperature dependence of the $\left(010\right)$ Bragg peak was followed as a function of temperature to quantify the critical behaviour of the sublattice magnetization in CoPS$_3$.  The results are shown in figure \ref{fig:beta}.  The Bragg peak intensity decreases as a power law with an exponent of $2\beta = 0.60\pm 0.01$ close to $T_N$, which was established to be $119.1 \pm 0.1$ K in the analysis.  The exponent was determined from fitting data in the temperature range $109 \le T < T_N$, or from a reduced temperature $\left(1-T/T_N\right) \le 0.085$.  The exponent is roughly similar to that expected for a phase transitions in a three-dimensional material, suggesting that CoPS$_3$ may be less two-dimensional than some of the other members of the MPS$_3$ family.  The temperature dependence is remarkably similar to that observed for NiPS$_3$, with the same exponent within approximately the same reduced temperature range.

% Figure 8
% end Figure 8

% Discussion
\section{\label{sec:Discussion}Discussion}
The present results largely confirm the previously published magnetic structure and properties of CoPS$_3$ \cite{Ouvrard82}, and can be compared with the other members of the MPS$_3$ family.

The magnetic structure and critical behaviour of the sublattice magnetization of CoPS$_3$ are very similar to NiPS$_3$ \cite{Wildes15}.  %The ordered moments for both compounds are also smaller than might be expected from fitting the paramagnetic susceptibilities, suggesting that spin dynamics and spin-orbit coupling may play an important role in these compounds.  
Conversely, CoPS$_3$ is shown to have XY-like anisotropy while the paramagnetic susceptibility of NiPS$_3$ is isotropic.  NiPS$_3$ was initially thought to be anisotropic, however, also with XY-like characteristics \cite{Joy}.   The apparent, previously observed, anisotropy in NiPS$_3$ appears to be due to the influence of gluing the sample to a support \cite{Wildes15}, possibly resulting in structural distortion through an effect like magnetostriction.  It may be that NiPS$_3$ is very close to being XY-like and would further resemble CoPS$_3$ if subjected to a controlled structural distortion.

CoPS$_3$ and NiPS$_3$ offer a strong contrast with FePS$_3$ and MnPS$_3$.  The in-plane antiferromagnetic order of FePS$_3$ \cite{Lancon} has, like CoPS$_3$ and NiPS$_3$, a ``zig-zag" structure.  However, the nature of its anisotropy is very different.  FePS$_3$ is a good example of an Ising-like system with the moments pointing normal to the planes.  The anisotropy in FePS$_3$ is very strong, creating an energy gap of $\sim 16$ meV in the spin-wave spectrum.  The magnetic structure of MnPS$_3$ is different, with each magnetic moment being antiferromagnetically coupled to its nearest neighbours and hence having no frustration.  The moments point almost normal to the planes, with a small in-plane component \cite{Ressouche}.  The anisotropy is small, producing a gap of $\sim 0.5$ meV, and appears to be due to a combination of dipole-dipole and single-ion anisotropies \cite{Wildes98}.

Understanding the nature of the anisotropies may be the key to understanding the differences amongst the magnetic properties of the MPX$_3$ compounds.  Dipole-dipole anisotropy will be present in all the compounds and would favour the moments pointing normal to the $ab$ planes \cite{Goossens}.  This is the dominant anisotropy in MnPS$_3$, and may explain the small out-of-plane components to the moments in CoPS$_3$ and NiPS$_3$.  A single-ion anisotropy may favour the moments to lie in the planes.  This would explain the moment orientation for CoPS$_3$ and NiPS$_3$, and the small tilt of the moments towards the $\mathbf{a}$ axis in MnPS$_3$ \cite{Ressouche}.  A small in-plane anisotropy also appears to be what drives the critical dynamics of the magnetic phase transition in MnPS$_3$.  The strong out-of-plane anisotropy in FePS$_3$ is clearly different and possibly makes this compound unique in the family.

An indication of the relative magnitudes of the magnetic exchange parameters in CoPS$_3$ and NiPS$_3$ results from a comparison with a calculated classical phase diagram for a two-dimensional honeycomb lattice \cite{Fouet}.  The comparison works well for MnPS$_3$ and FePS$_3$, where the magnetic exchange parameters, derived from neutron inelastic scattering experiments \cite{Lancon, Wildes98}, show that the observed magnetic structure is stable.  The phase diagram has also been successfully used in a Monte-Carlo study of Na$_2$Co$_2$TeO$_6$, which is another cobalt-containing low-dimensional compound with a very similar zig-zag magnetic structure \cite{Lefrancois}.  Exchange parameters up to the third-nearest neighbour in the $ab$ planes needed to be included to fit the magnon dispersion for MnPS$_3$ and FePS$_3$, although for both compounds the exchange between second-nearest neighbours was practically zero.  This should not be the case for CoPS$_3$ and NiPS$_3$.  Assuming that the nearest-neighbour exchange is antiferromagnetic, which is reasonable considering the sign of the Weiss temperatures, the magnetic structures for these compounds are stable under the following conditions:
\begin{equation}
\begin{array}{l}
  2J_2 \ge J_1 \\
  4J_3 \ge 2J_2+J_1
\end{array}
\end{equation}
where $J_n$ are the exchange parameters between the $n^{\mbox{th}}$-nearest neighbours and all $J$ have the same sign.  The bounds suggest that the exchange between second-nearest neighbours should be relatively large in CoPS$_3$ and NiPS$_3$, and that all three exchange parameters should be antiferromagnetic.

Further experiments examining the spin dynamics and properties of the phase transitions in these compounds are planned.

% Conclusion
\section{\label{sec:Conclusion}Conclusion}
CoPS$_3$ is shown to be an antiferromagnet with apparent XY-like anisotropy.  The previously published magnetic structure was largely correct, although there appears to be a small out-of-plane component to the ordered moments. The ordered moment is consistent with having a small orbital contribution to the total angular momentum of the Co$^{2+}$ ion. The magnetic structure and critical properties are very similar to those of NiPS$_3$, a sister compound, although the latter appears to be more isotropic in its paramagnetic susceptibility.

%Acknowledgements
\section{Acknowledgements}
The authors thank the Institut Laue-Langevin and the CEA for the use of the neutron-beam instruments, and the CNRS for the use of the magnetometers.  Thanks also to E. Eyraud from Institut N{\'e}el for the technical support in the SQUID magnetometry measurements.  ARW thanks Dr. Oscar Fabelo for his assistance in using the FullPROF package.  Figure \ref{fig:CoPSMag} was created with the VESTA software package \cite{VESTA}.

% References

\end{document}